\documentclass[useAMS,usenatbib]{mn2e}
\usepackage{graphicx, amsmath, amssymb,epsfig}

\def\kms{\mbox{km~s$^{-1}$}}
\def\kpc{\mbox{kpc}}

\def\LCDM{\mbox{$\Lambda$CDM}}

\def\Mpch{\mbox{$h^{-1}$Mpc}}
\def\Mvir{\mbox{$M_{\rm vir}$}}
\def\M200{\mbox{$M_{\rm 200}$}}

\def\Msun{\mbox{$M_\odot$}}
\def\Msunh{\mbox{$h^{-1}M_\odot$}}

\def\R200{\mbox{$R_{\rm 200}$}}

\def\Vmax{\mbox{$V_{\rm max}$}}
\def\V200{\mbox{$V_{\rm 200}$}}

\newcommand{\HI}{HI}

\newcommand{\apjs}{ApJ Suppl.}
\newcommand{\apj}{ApJ}
\newcommand{\apjl}{ApJLett}
\newcommand{\aj}{AJ}

\newcommand{\mnras}{MNRAS}

\title[Velocity Function]{Abundance of Field Galaxies.}  \author[Klypin, et
al.]{Anatoly~Klypin$^{1}$\thanks{E-mail: aklypin@nmsu.edu},
  Igor~Karachentsev$^{2}$, Dmitry Makarov$^{2}$,
  and Olga~Nasonova$^2$\\
  \vspace{-0.2cm}\\
  $^{1}$New Mexico State University, Las Cruces, NM 88001, USA\\
  $^{2}$Special Astrophysical Observatory, Nizhny Arkhyz, Russia\\
}
\begin{document}

\date{Draft, 2014, May 12}

\pagerange{\pageref{firstpage}--\pageref{lastpage}} \pubyear{2002}

\maketitle

\label{firstpage}

\begin{abstract}

  We present new measurements of the abundance of galaxies with a
  given circular velocity in the Local Volume: a region centered on
  the Milky Way Galaxy and extending to distance $\sim$10\,Mpc. The
  sample of $\sim 750$ mostly dwarf galaxies provides a unique opportunity to study the
  abundance and properties of galaxies down to absolute
  magnitudes $M_B\approx -10$, and virial masses $M_{\rm vir}=
  10^{9}\Msun$. We find that the standard \LCDM\ model gives
  remarkably accurate estimates for the velocity function of galaxies
  with circular velocities $V\gtrsim 70\,\kms$ and corresponding
  virial masses $M_{\rm vir}\gtrsim 5\times 10^{10}\Msun$, but it
  badly fails by over-predicting $\sim 5$ times the abundance of large
  dwarfs with velocities $V= 30-40\,\kms$. The Warm Dark Matter (WDM)
  models cannot explain the data either, regardless of mass of WDM
  particle.  Just as in previous observational studies, we find a
  shallow asymptotic slope $dN/d\log V \propto V^{\alpha}, \alpha
  \approx -1$ of the velocity function, which is inconsistent with the
  standard \LCDM\ model that predicts the slope $\alpha =-3$. Though
  reminiscent to the known overabundance of satellites problem, the
  overabundance of field galaxies is a much more difficult
  problem. For the standard \LCDM\ model to survive, in the 10~Mpc
  radius of the Milky Way there should be $1000$ not yet detected galaxies with
  virial mass $M_{\rm vir}\approx 10^{10}\Msun$, extremely
  low surface brightness and no detectable \HI~ gas. So far none of
  this type of galaxies have been discovered.

\end{abstract}

\begin{keywords}
cosmology: theory -- dark matter -- galaxies: halos -- methods: N-body simulations.
\end{keywords}

\section{Introduction}

The velocity function, which is defined as the abundance of galaxies
with a given circular velocity, is one of fundamental statistical
properties of galaxies. It is a kin of the much more well known and
studied statistics: the luminosity function - the abundance of
galaxies with a given luminosity. The luminosity function is easier to
measure, and indeed there are numerous estimates of the luminosity
function
\citep[e.g.,][]{Norberg2002,Bell2003,Blanton2005,Montero-Dorta2009}. From
the theoretical point of view there are substantial differences
between luminosity and velocity functions. It is much more difficult
to make theoretical predictions for the luminosities. Galaxy
luminosity and stellar mass are the results of a complicated history
of star formation of a galaxy. It depends on accretion and merging
history and also on many other processes, which operate when a galaxy
evolves -- the stellar winds, supernovae explosions, galactic
fountains -- to name a few.  This makes the luminosity and stellar
mass very valuable tools to study the evolution of the Universe, but
it makes them very difficult to predict.

Testing the theoretical predictions of the abundance of galaxies can
done using the Semi-Analytical Models (SAMs)
\citep[e.g.,][]{White1991,Somerville1999,Benson2003,Somerville2012}.
Unfortunately, SAMs use many assumptions and parameters, which make
theoretical predictions somewhat uncertain. The main source of the
uncertainty is due to the lack of detailed understanding of how
galaxies form and evolve in the cosmological framework. Another
popular way of relating dark matter halos with galaxies are the halo
abundance matching (HAM) \citep{Kravtsov2004,Conroy06,Trujillo2011}
and the halo occupation distribution (HOD)
\citep{Berlind2002,Kravtsov2004,Zentner2005}.  These methods are often
used to predict galaxy clustering at different redshifts. However,
they {\it assume} galaxy luminosity or stellar mass functions, and
thus cannot be used to test theory when it comes to predictions of
abundances of galaxies.

Because the circular velocity measures the mass in the inner region of
a dark matter halo (e.g., about $\sim 20\,\kpc$ for a Milky-Way mass
halo) where the observed galaxy is situated, and because the circular
velocity does not depend on the complicated history of star formation,
the velocity function of galaxies can be predicted much more
accurately than the luminosity function. This makes the circular
velocity function a useful tool for testing the theory
\citep{Cole1989,Shimasaku1993,Gonzalez2000, Zavala2009,Trujillo2011}.

The abundance of satellites of the Local Group is an example of the
power of the velocity function as a test for cosmological predictions
\citep{Klypin1999,Moore1999}. Comparison of the predicted abundance of
subhalos with given circular velocity in cosmological $N$-body
simulations with the observed number of satellites clearly indicate a
large disparity between the theory and observations. There are
explanations for the disagreement, which include a variety of
different effects, including photoheating during the reionization epoch
\citep{Barkana1999,Bullock2001,Shapiro2004} and  stellar
feedback \citep{Dekel1986,MacLow1999,Kravtsov2010}.  Modifications of
the cosmological model are also used to address the problem.
Suppression of the spectrum of fluctuations on small scales in models
of the warm dark matter  results in substantial reduction of
predicted small halos, which is the reason why WDM models are often
used as explanation for the overabundance of the subhalos
\citep{Colin2000,Kamionkowski2000,Bode2001,Kennedy2013,Polisensky2014}.
See, however, \citet{Schneider2014,Schultz2014}.

Velocity function of galaxies addresses some of the same key issues as
the abundance of satellites in the Local Group (e.g., are there too
many dwarf halos predicted by the \LCDM~ model). However, in many
respects these are a different statistics. Velocity function measures the abundance
of all galaxies -- not only the satellites. It may seem
couter-intuitive, but for a given cut of the circular velocity, most
of the objects are ``parents'': galaxies or DM halos in simulations
that do not belong to a larger galaxy or halo
\citep{Klypin2011,Nuza2013,Guo2014}. Another difference is the
fraction of different morphological types. Most of dwarf galaxies in
the Local Group are dwarf spheroidal galaxies, whereas most of the
galaxies in the Local Volume  are star-forming dwarf irregular galaxies.

The velocity function is relatively easy to predict theoretically
\citep{Klypin2011,Trujillo2011}, but much more difficult to measure in
observations.  So far there were some attempts to produce
observational estimates using SDSS data \citep{Gonzalez2000,Sheth2003,
  Choi2007,Chae2010}, HIPASS \citep{Zwaan2010} and ALFALFA
\citep{Papastergis2011}. In spite of the progress in the measurements,
there are some disagreements. For example, SDSS
\citep{Choi2007,Chae2010} and HIPASS \citep{Zwaan2010} data indicate
that the VF becomes constant at velocities $V\lesssim 100\,\kms$,
while \citet{Papastergis2011} find that VF keeps increasing even at
very small velocities $dN/d\log(V) \propto V^\alpha$ with the slope
$\alpha\approx -0.85$. Galaxies in both the HIPASS and the ALFALFA are
selected by \HI~ fluxes, which means that they miss early-type galaxies.

While predicting circular velocity is easier than that of the stellar
mass or luminosity, it still requires some effort and needs careful
estimates and corrections due to different effects. There are
different steps toward accurate predictions of the velocity
function. The first step is large cosmological N-body simulations with
high mass and force resolution. Resolution is an important
factor. Because the maximum of the circular velocity is reached at a
small fraction the virial radius, the resolution of simulations needed
for accurate estimates of the velocity function is typically 5-10
times better than that needed for the halo mass function
\citep{Klypin2013}. Only recently simulations with this resolution and
large volume became available providing us with needed estimates.  In
addition to high-quality simulations one needs to make corrections due
to baryons: gas and stars in central regions of galaxies make the
circular velocity larger
\citep[e.g.,][]{Mo1998,Klypin02,Dutton2011,Trujillo2011}. These
corrections are small for galaxies with circular velocities below
$\sim 100\,\kms$, which are dark matter dominated even in central
$5-10\,\kpc$ regions. For larger galaxies the corrections can be as
large as 20-50\%.

 The paper is organized as follows. In Section 2 we present our
 observational sample.  Theoretical estimates are presented in Section
 3. Results are presented in Section 4.  Discussion is given in
 Section 5.

\section{Galaxies in the Local Volume: observational sample}

\subsection{Description of the sample}

Volume limited sample of galaxies within the Local Volume were
substantially extended and improved over the last decade
\citep{Karachentsev2004,Kar07ap,Karachentsev2013}.  Evolved from
the original  list of 179 galaxies \citep{Kraan1979} the current version of 
the Updated Nearby Galaxy Catalog \citep{Karachentsev2013} contains 869
galaxies with redshift-independent distances $D<11$\,Mpc or radial
velocities with respect to centroid of the Local Group
$V_\textrm{LG}<600$\,\kms.  The sample was updated by results of a systematic
search for new low surface brightness (LSB) galaxies and follow up
radio- and optical observations.  Significant number of new irregular
dwarf galaxies were added by blind \HI{} surveys such as HIPASS and
ALFALFA.  The  redshifts surveys such as SDSS, 2dF and 6dF improved
our knowledge not only of distant Universe, but also for the Local
Volume.  Special surveys for extremely low surface brightness
satellites around Milky Way, Andromeda and M\,81 reveal the galaxies
with total luminosity about $M_V\sim-4$.

The redshift is not reliable distance indicator in the Local Volume
because of peculiar velocities.  For instance, observed 70--100\,\kms{}
virial line-of-sight velocities of galaxies in the nearby groups
are comparable with the recession velocity of the groups $\sim
300\,\kms$~ \citep{Karachentsev2005}.  Fortunately, because of its
proximity, redshift-independent distances have been measured for most of
the Local Volume galaxies.  A large fraction of objects, namely 311,
have distance estimations with high accuracy of 5--10\,\%, which are
based on the tip of the red giant branch or cepheids methods.  Most of
such galaxies lie below 5\,Mpc \citep[][see Fig.~3]{Karachentsev2013}.

In the current work we test 3 subsamples of the Local Volume galaxies.
The subsample with distances $D\leq 10$\,Mpc contains 733 galaxies, of
which 652 objects are brighter than $M_B=-10$ and 426 are brighter
than $M_B=-13$.  The $D\le8$\,Mpc subsample consists of 568 objects,
where 488 and 298 are brighter than $M_B=-10$ and $M_B=-13$,
respectively.  The $D\leq 6$\,Mpc set comprises 378 objects, 303 and 170
are brighter than $M_B=-10$ and $M_B=-13$, correspondingly.

The Local Volume catalog has a significant fraction of early type
galaxies. This is very important when observational results on the
velocity function are compared with theoretical expectations: galaxies
of all morphological types are counted in the Local Volume catalog.
This makes a significant difference with galaxy catalogs such as
HIPASS \citep{Zwaan2010} and ALFALFA \citep{Papastergis2011}, which
are based on \HI~ observations, and thus miss gas poor galaxies.

The fraction of early type galaxies in the Local Volume increases with
the decreasing luminosity.  Only 6--7\,\% of bright ($M_B<-16$) objects
are lenticular or elliptical galaxies, while the fraction of spheroidal
among of all dwarfs with $-10> M_B > -13$ is 31\,\%.

Galaxies that do not have the \HI~ velocities are mostly early types
(E's or dSph, $\sim 10$\% of all galaxies). The rest are predominately
dwarf galaxies, for which \HI~ measurements are not yet available
($\sim 10$\% of all galaxies).
 For the galaxies without \HI~
line-width measurements, we assign the line-of-sight rms velocities
$V_{\rm los}$ using the average luminosity-velocity (L-V) relation in
the K-band for galaxies with the measured line-width. In order to
construct the relation we use following data.

\begin{figure}
\centering
\includegraphics[width=0.48\textwidth]{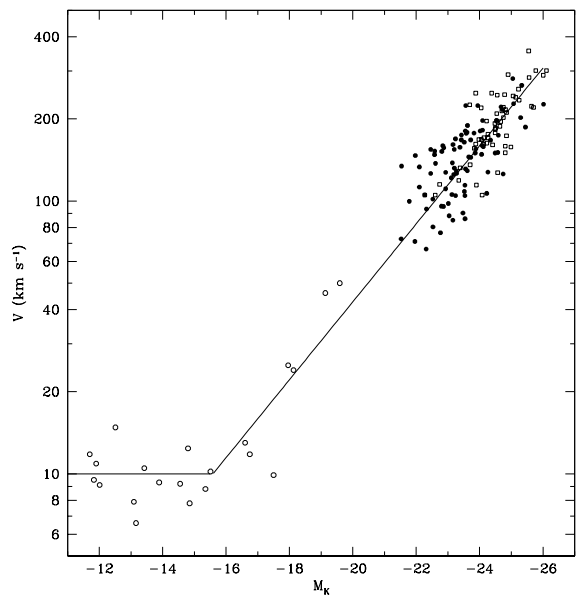}
\caption{Dependence of stellar velocity dispersion on the  K-band magnitude for early-type
  galaxies in different observational samples. Open circles are for
  dwarf galaxies in the Local Group.  Galaxies in the ATLAS3D catalog
  \citep{Cappellari2013} are shown as filled circles. Open squares are
  for a compilation of early-type galaxies in \citet{Trujillo2011}. The
  lines in the plot show approximation eq.~(\ref{eq:FJ}) for $V_{\rm los}(M_K)$. }
\label{fig:FJ} 
\end{figure}

For dwarf galaxies we use stellar velocity dispersions of the Local
Group dwarf spheroidals and  ellipticals given by
\citet{Kirby2014} and \citet{Geha2010}. K-band magnitudes for those
galaxies are taken from \citet{Karachentsev2013}.  For larger galaxies
we use stellar velocity dispersions in ATLAS3D catalog for early type
galaxies \citep{Cappellari2013}. We cross-correlate the ATLAS3D with
the 2MASS catalog \citep{Huchra2012} and identify galaxies that are
listed in both catalogs. In order to reduce errors in distances (and
thus luminosities), we  use only galaxies with distances larger than
16~Mpc. We  use also circular velocities for early type galaxies given
in Appendix of \citet{Trujillo2011}. For these galaxies we assume
$(K-B)=-3.5$ color correction and divide circular velocities by
$\sqrt{3}$ to estimate the line-of-sight rms velocities. 
Figure~\ref{fig:FJ} presents the results.

The plot shows that for bright galaxies with $M_K<-18$ the line-width
$V_{\rm los}$ depends on the luminosity, but this is not the case for
dwarfs. Both effects are well known: a Faber-Jackson-type relation for
bright ellipticals and the lack of dependence of dynamical mass within
central $\sim 500$\,pc for dwarf spheroidals
\citep[e.g.,][]{Strigari2008}. As one may have expected, the spread of
the $V_{\rm los}$-L relation is relatively large: about 20\% for $V_{\rm los}$
at given $M_K$. Nevertheless, with this accuracy, results in
Figure~\ref{fig:FJ} give us a way to estimate line-width for galaxies
that we do not have \HI~ measurements for. Specifically, we use the
following approximation, which provides a fit to the observational
data:
\begin{equation}
V_{\rm los} = \left\{ \begin{array}{ll}
         70\cdot 10^{-(21.5+M_K)/7}\,\mbox{km\,s}^{-1} & \mbox{if $M_K<-15.5$},\\
         10\, \mbox{km\,s}^{-1} & \mbox{if $M_K>-15.5$}.\end{array}  
\right. \label{eq:FJ}
\end{equation}

\begin{figure}
\centering
\includegraphics[width=0.48\textwidth]{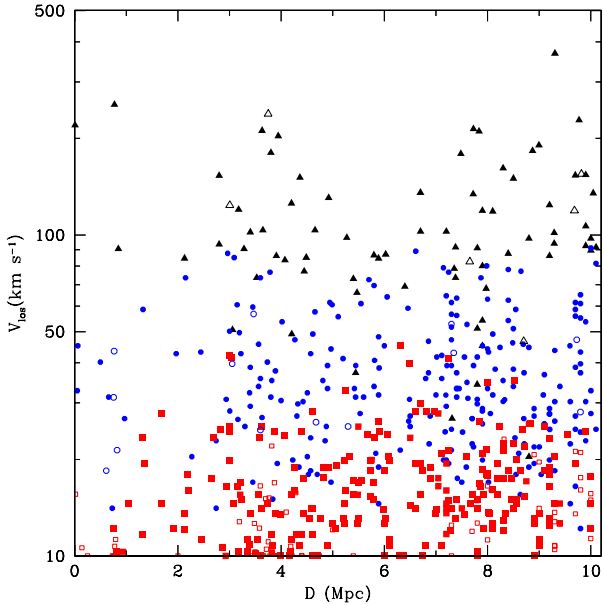}
\caption{Distribution of line-widths $V_{\rm los}$ of galaxies in
  observations as the function of distance from the Milky Way. Empty
  (filled) circles are for early (late) type galaxies. Colors code
  bright (black, $M_B>-18$), intermediate (blue , $-14>M_B>-18$), and
  dwarf (red, $M_B>-14$) galaxies. The enhancement of the number of
  galaxies at the distance $D\approx 3.5-4\,$Mpc is due to large
  groups with central galaxies NGC5128, M81, and IC342. }
\label{fig:VDdiagrams}
\end{figure}

Low accuracy of the $b/a$ for some small galaxies force us to use
the distribution of the line-widths $V_{\rm los}$ not corrected for
the inclination. In this respect we follow the suggestion of
\citet{Papastergis2011}, and use line-widths as the main
characteristics of observed galaxies. In order to simplify the comparison
with the theory, instead of the full width $W_{50}$ we use half-width
$V_{\rm los}$ as a proxy for the projected circular velocity.

Figure~\ref{fig:VDdiagrams} presents the distribution of line-widths
$V_{\rm los}$ for galaxies in the Local Volume. The overall increase
of the number of galaxies at large distances is simply the reflection
of increasing volume. However, there is a real drop in the
number-density of galaxies at distance $D\approx 2\,$Mpc, which is
followed by an enhancement at $D\approx 3.5\,$Mpc due to large galaxy
groups at that distance. Another effect is a gradual increase in the
number of early-type galaxies for low-luminosity galaxies. Incompleteness of the sample
 manifests itself as the apparent lack of galaxies with $V_{\rm los}\lesssim 15\,\kms$ at $D>5\,$Mpc.

There are two ways of comparing the observational results of the
distribution of line-width with the theory: (1) One can apply
corrections to the theoretical predictions as was suggested by
\citet{Papastergis2011}. This is done separately for disk galaxies and
for early-type galaxies.  For disk galaxies we assume a random
orientation of disks in space, but no correction is applied for
elliptical galaxies, for which we assume $V_{\rm los}=V/\sqrt{2}$.  (2)
One can also de-project the observational sample by assuming a random
orientation of disk galaxies. This can be done in a number of
ways,  here we use a parametric fitting. We use an analytical function
of the distribution of the circular velocities with free
parameters, then parameters are tuned to reproduce the observed
distribution of line-width.

\begin{figure}
\centering
\includegraphics[width=0.48\textwidth]{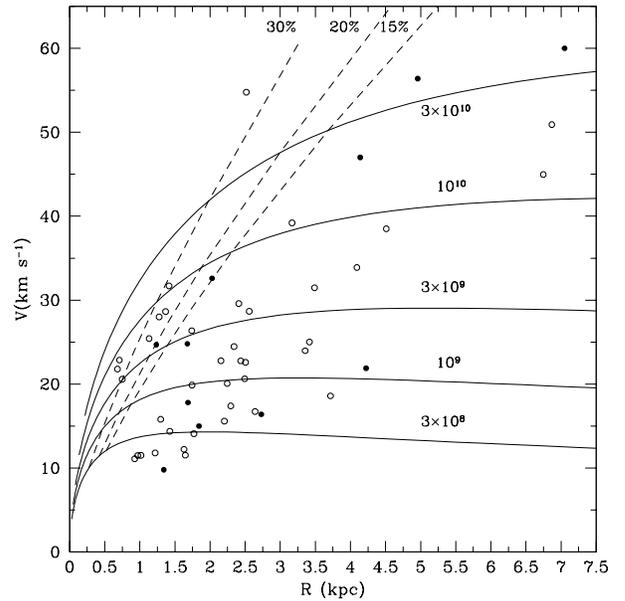}
\caption{ Extent $R$ of neutral hydrogen in galaxies with different
  rotational velocities $V$. Open and filled circles show observed
  galaxies in the samples of \citet{Begum2008} and
  \citet{Moiseev2014}, correspondingly. Full curves present circular
  velocity profiles for dark matter halos with an average
  concentration and virial mass indicated in the plot.
 Dashed curves
  show errors in $\Vmax$ due to final extent of \HI~ gas: the
  correction is less than the indicated value in the plot for all
  galaxies to the right of a corresponding dashed curve. 
Though for
  some galaxies the error may be 20-30\%, the plot demonstrates that
  \HI~ typically extends far enough to measure $\Vmax$ with relatively
  small error. }
\label{fig:GasExtent}
\end{figure}

Because neutral hydrogen typically extends well beyond the optical
radius even for dwarf galaxies
\citep[e.g.,][]{Cote2000,Swaters2002,Begum2008,Walter2008,Moiseev2014},
measurements of \HI~ linewidths using $W_{50}$ provide estimates of
the the rotation velocities at very large distances. To be more
specific, they provide the {\it maximum} rotational velocity in the
region where there is a detectable amount of neutral hydrogen. But is
this region large enough? When we compare these velocities with
theoretical predictions, we use $\Vmax$ values of dark matter halos,
which also occur at large distances. Therefore, we need to find
whether the neutral hydrogen extends far enough to probe $\Vmax$.

For most of the galaxies in our sample the extent of the \HI~
component is unknown. Still there are many galaxies for which the
measurements exist and we will use two samples of galaxies to shed
light on the issue. The first sample is the FIGGS survey
\citep{Begum2008}, which, among other parameters provides $W_{50}$ and
the \HI~radius that corresponds to a column density of $10^{19}$~atoms
per cm$^{-2}$. We select galaxies that have distance measurements
obtained with the Tip of the Red Giant Branch (TRGB) method presented
in \citet{Rizzi2007}.  In Figure~\ref{fig:GasExtent} open symbols show
FIGGS rotational velocities corrected for inclination. In addition we
show a compilation of galaxies given by \citet{Moiseev2014} with
  measured \HI~ extent as filled circles in the same Figure.
Similar results were found by  \citet{Ferrero} and by \citet{Papastergis2014}.

The full curves in the plot show circular velocity profiles for dark
matter halos with a NFW density distribution and an average
concentration. The virial masses for each halo are given in the plot in units of solar
mass.  Dashed curves show potential correction to $\Vmax$ values due
to the fact that \HI~ is measured only in the central halo region: the
correction is less than the indicated value in the plot for all galaxies
to the right of a  corresponding dashed curve.  If we assume that every
galaxy has a NFW profile with the average concentration, than there should
be a NFW curve passing through every galaxy in the plot. For example,
if the galaxy is to the right of a dashed curve with a 15\% label, the
$W_{50}$ (with all the corrections to the inclination) gives a $\Vmax$
with less than 15\% error.

In spite of the fact that for some galaxies correction due to the final
extent of \HI~ gas may be $\sim (20-30)$\%, the plot shows that \HI~
typically extends far enough to measure $\Vmax$ with relatively small
error.

\subsection{Luminosity Function and Completeness of the sample}
\begin{figure}
\centering
\includegraphics[width=0.48\textwidth]{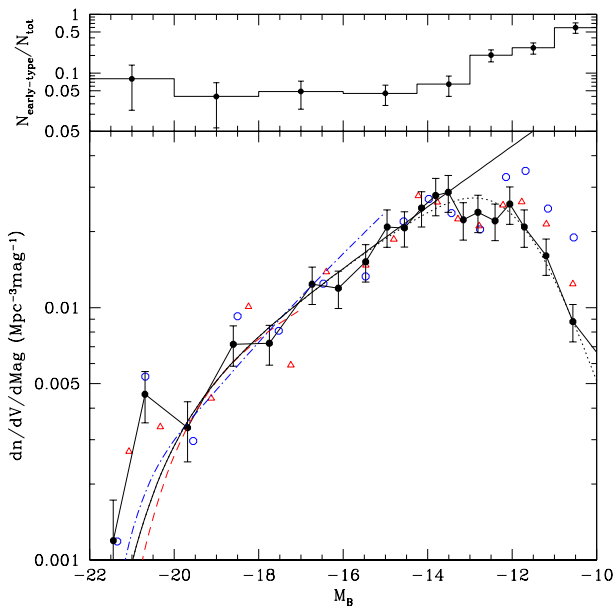}
\caption{{\it Bottom:} Luminosity function of galaxies in the
  Local Volume.  The filled circles are for the 10\,Mpc sample. 
  Error bars show Poissonian fluctuations.  The luminosity function
  for the 6\,Mpc and 8\,Mpc samples are presented with open circles
  and triangles. For comparison the luminosity functions for the SDSS
  (dot-dashed curve \citep{Blanton2005}) and the 2dFGRS (dashed curve
  \citep{Norberg2002}) are also shown.  The data indicate that the
  Local Volume function is complete for $M_B<-14$. The full curve
  shows the Schechter approximation with the slope $\alpha=1.30$ and
  $M_*=-20.0+5\log(h)$.  At smaller magnitudes the observed luminosity
  function bends down indicating that the sample is less complete. The
  dotted curve shows a fit for the luminosity function in the 10~Mpc
  sample in the $M_B=-12-14$ range.  {\it Top:} Fraction of early-type
  galaxies in the 10~Mpc sample. The fraction is almost constant $\sim
  10\%$ for galaxies brighter than $M_B=-13$. It steeply increases for
  smaller galaxies mostly due to dSph satellites around bright
  galaxies. }
\label{fig:LumFun}
\end{figure}

The sample of galaxies in the Local Volume was gradually improved and
extended over years.  \citet[][Section 4]{Karachentsev2004} discuss
completeness of the earlier sample and conclude that within 8~Mpc
radius the sample was 70-80 percent complete, implying that about 100
galaxies were missed in that sample. \citet{Tikhonov09} studied
completeness of the Local Volume using two methods. They  used the updated
sample, which had $\sim 100$ more galaxies, and thus it was nearly
complete within 8~Mpc radius for galaxies with $M_B<-12$.  In
both methods the ratio of the number of dwarf galaxies to the number
of bright galaxies was used as an indicator of completeness since the
ratio should not depend on the distance.  First, the number of bright
galaxies ($M_B<-15$) and the number of dwarf galaxies ($M_B=-12-14.5$)
inside radial shells of 1~Mpc width were found. If the sample is not
complete, we would expect a decline with the distance of the number of
dwarf galaxies. The ratio of the number of dwarf to large galaxies did
not indicate any decline and confirm the completeness of the
sample. Second, galaxies in the zone of avoidance were counted and
compared with the counts in the direction of the galactic pole. For
the same two subsamples ($M_B<-15$ and $M_B=-12-14.5$)
\citet{Tikhonov09} found 28 bright galaxies and 18 dwarfs close to the
galactic plane ($|b|<15^o$). In the direction of the galactic pole
($|b|>75^o$) they found 28 giants and 16 dwarfs. This gives the ratio
of dwarfs/bright galaxies equal to 0.64 in the the direction of the
galactic pole and 0.57 in the galactic plane. Again, results are
compatible with the completeness of the sample used at that time.  The
present sample is nearly complete to distances $D<10\,$Mpc. This
almost doubles the volume of the sample as compared with what was used
by \citet{Tikhonov09}.

Figure~\ref{fig:LumFun} shows the luminosity function of galaxies in
the \citet{Karachentsev2013} catalog for different subvolumes.
Results for the 6~Mpc and 8~Mpc samples were normalized to have the same
number-density of galaxies brighter than $M_B=-14$ as in the 10~Mpc
sample. There are some variations between different subsamples, but
for galaxies brighter than $M_B=-14$ these variations are consistent
with pure shot-noise. At smaller luminosities
there are clear indications of incompleteness with smaller volumes
having more dwarf galaxies with $M_B=-10-12$.

For comparison, Figure~\ref{fig:LumFun} also presents the luminosity
function in the 2dFGRS galaxy catalog \citep{Norberg2002} and the SDSS
sample \citep{Blanton2005}.  The 2dFGRS luminosity function was given
in $b_J$ magnitudes. We scaled it to the B-magnitudes using the
relation $b_j=B-0.28(B-V)$ \citep{Norberg2002} and taking $B-V
=0.5$. The 2dFGRS luminosity function was estimated only for
relatively bright galaxies with $M_B<-17.2$.  The SDSS luminosity
function extends to significantly smaller galaxies with $M_B\approx
-15$ because it was based on a shallow SDSS subsample for galaxies
with distances in the range $10-150 h^{-1}\,$Mpc.  We use the double
Schechter ``corrected'' approximation in Table~3 of
\citet{Blanton2005} for the g-band magnitudes, which we convert to
B-magnitudes using the relation $g= B -0.235-0.34[B-V-0.58]$
\citep{Blanton2007} and taking $B-V =0.5$.

The full curve in Figure~\ref{fig:LumFun} presents a Schechter fit to
the LV data with $M_B<-14$:

\begin{equation}
 \Phi(L)dL = \phi_*\left(\frac{L}{L_*}\right)^\alpha 
              \exp\left(-\frac{L}{L_*}\right)\frac{dL}{L_*},
\label{eq:Schecter}
\end{equation} 
where $ \phi_* = 1.25\times 10^{-2}h^3{\rm Mpc}^{-3}$, $\alpha=1.3$, and
$M_{*,B}=-20.0+5\log(h)$.  Comparison with the SDSS and 2dFGRS
luminosity functions indicates that the Local Volume is a typical
sample of galaxies for the volume probed. The only systematic
deviation which we find is an excess in the Local Volume of very
bright galaxies with $M_B\approx -21$. Otherwise, it is a normal sample.

We model the incompleteness of the sample at $M_B> -14$ by dividing
the measured luminosity function (dotted curve in
Figure~\ref{fig:LumFun}) by the Schechter approximation extrapolated
from the brighter galaxies. The ratio $f_{\rm select}=N_{\rm
  obs}/N_{\rm Sch}$ of the two gives the selection function of
galaxies in the Local Volume. It can be approximated as:

\begin{equation}
f^{-1}_{\rm select}(M_B)= 1 + 10^{0.6(M_B-M_0)}, \quad M_0 =-11.9.
\label{eq:missed}
\end{equation} 
According to these results, the 90 percent completeness is at $M_B
=-13.5$, which on average corresponds to $V_{\rm los} \approx 20\,\kms$.
We estimate that the sample misses 1/2 of galaxies at $M_B =-12$ and
$V_{\rm los} \approx 13\,\kms$.  Motivated by these results, we estimate
the selection function in velocities $V_{\rm los}$:
\begin{equation}
f^{-1}_{\rm select}(V_{\rm los})= 1 + \left[\frac{V_{\rm los}}{13\,\kms}  \right]^{-4.5}.
\label{eq:correct}
\end{equation} 
We make correction for incompleteness of the catalog by multiplying
the observed number of galaxies with given line-width $V_{\rm los}$ by
$f^{-1}_{\rm select}(V_{\rm los})$ given by eq.~(\ref{eq:correct}). This
correction plays a role only for very small galaxies. For example, the
correction is only 5 percent for galaxies with $V_{\rm los} =25\,\kms$, and
is totally negligible for larger galaxies.

\begin{figure}
\centering
\includegraphics[width=0.48\textwidth]{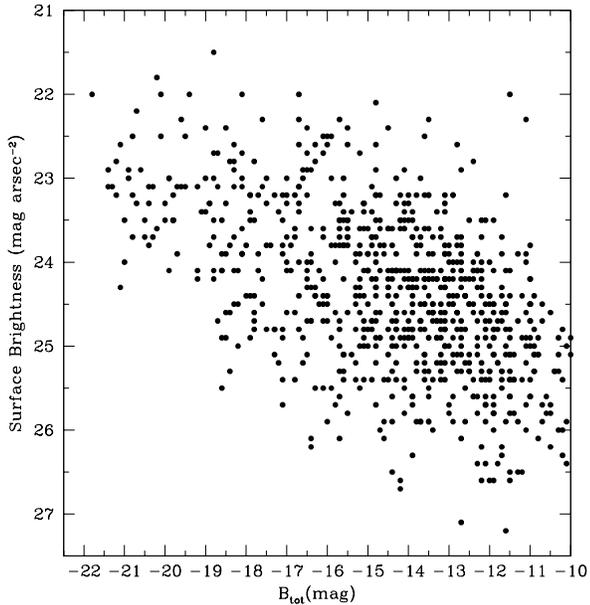}
\caption{ Average surface brightness of galaxies in the UNGC catalog as
  function of total absolute magnitude in B band.  Note a very large
  range of surface brightnesses in the catalog: dwarf galaxies can be
  $\sim 4$~magnitudes less bright than giant galaxies. With large
spread the surface brightness follows the luminosity. }
\label{fig:Surface}
\end{figure}

One may wonder if the surface brightness (SB)
completeness could be the cause for the disagreement between the
$\LCDM$ model and observations. The SB completeness of the sample has
already been discussed in \citet{Karachentsev2013}. Here we reproduce
some of the results and arguments of \citet{Karachentsev2013}.

Figure~\ref{fig:Surface} shows the average surface brightness within
the Holmberg radius of galaxies in the UNGC catalog (see also Figure
6 in \citet{Karachentsev2013}).  There is a large spread of SB for a
given total absolute magnitude in B band, $B_{tot}$. There is also a
clear trend: less bright galaxies have on average lower SB. The data
in the Figure are consistent with a simple assumption that with large
spread the surface brightness follows the luminosity. There is an
indication that the sample becomes less complete below SB$\approx
26\,{\rm mag}\,{\rm arcsec}^{-2}$, but it is not clear whether this is a
separate limitation of the sample or just a manifestation of
incompleteness below $B_{\rm tot}\approx -12$.

\section{Theoretical predictions for velocity function}

There are a number of steps, which we make in this paper in order to predict the
distribution of line-widths $V_{\rm los}$ of galaxies for a cosmological
model:
\begin{itemize}
\item Find the theoretical distribution of circular velocities $V$ of dark
  matter halos. Large cosmological simulations often provide those
  \citep[e.g.,][]{Gonzalez2000,Klypin2011,Nuza2013,Schneider2014}.
  The distribution function $dN/dV$ must include subhalos. If it does not, it
  should be corrected.
\item Correct the dark matter circular velocities for the effect of baryonic infall.
\item Assuming a random orientation of galactic disks, and, using the
  observed fraction of early-type galaxies, make a prediction for the
  distribution of line-widths $dN/dV_{\rm los}$.
\end{itemize}

Our predictions rely on a number of assumptions. We assume that
maximum of the circular velocity of a dark matter halo measured in
cosmological simulations and corrected for baryonic infall gives an
estimate of the HI line width. This is a reasonable assumption, if
neutral hydrogen extends to large radii where dark matter circular
velocity reaches its maxiumum. As Figure~\ref{fig:GasExtent} shows,
this seems to be the case as indicated by those galaxies that have
detailed measurements of HI rotation curves. Our procedure also
implies that processes related to star formation and stellar
feedback do not change the total mass of galaxies (including the dark
matter mass) inside the radius of HI extent (typically $\sim 3$ times
the optical radius of galaxies). More detaled discussion of these
effects is presented in Section~6.

We use the MultiDark suit of simulations \citep{Klypin2014} to construct
the velocity function $dN/d\log V$ in the $\LCDM$~model. Specifically,
we use the Bolshoi \citep{Klypin2011} and MultiDark \citet{Prada2012}
simulations for WMAP7 cosmological parameters. The the Planck
cosmology we use BolshoiP and MDPL simulations \citep{Klypin2014}. 
These simulations are done with the ART and Gadget  codes.

Halos in the simulations were identified with the Bound Density
Maximum (BDM) spherical overdensity code \citep{Riebe2013}. For each
halo or subhalo the halofinder provides the maximum circular velocity
$\Vmax$. In the following instead of $\Vmax$ we use notation $V$ and
call it the circular velocity.

\begin{figure}
\centering
\includegraphics[width=0.48\textwidth]{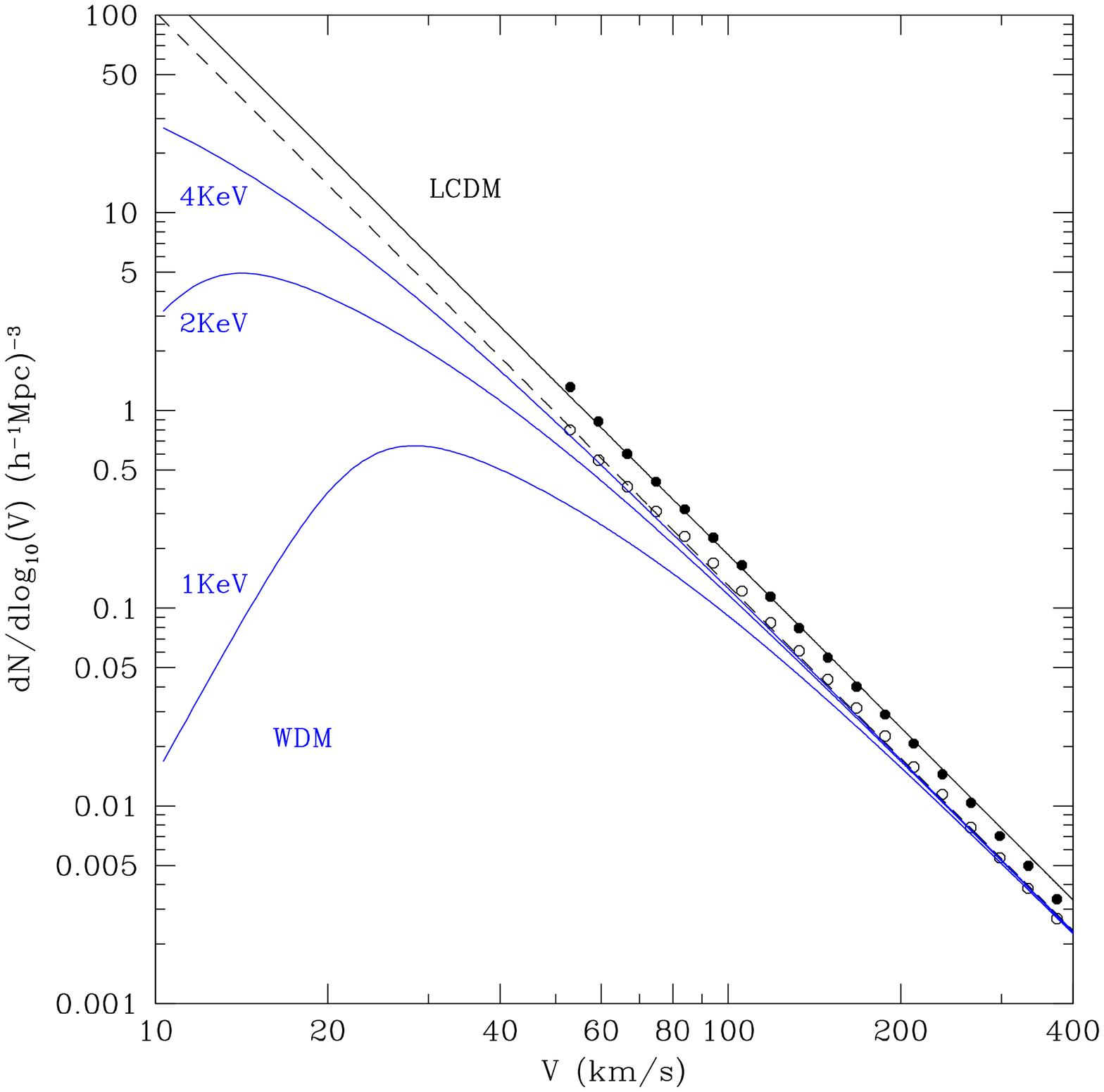}
\caption{Circular velocity function for halos in the $\LCDM$ and WDM
  models. Open circles are results from the Bolshoi simulation
  \citep{Klypin2011} for WMAP7 cosmology. The dashed line is the
  power-law approximation. Filled circles are for the BolshoiP
  simulation \citep{Klypin2014} for the Planck cosmology. The top full
  line shows a power-law fit for this cosmology. Other curves are
  analytical fits for the WDM model with WMAP7 cosmological parameters
  \citep{Schneider2014} with different neutrino mass indicated in the
  plot.}
\label{fig:LCDM}
\end{figure}

The Bolshoi and BolshoiP simulations are complete for halos and
subhalos down to $V=50\,\kms$ \citep{Klypin2011}. Multidark and MDPL
simulations are complete down to $\sim 160\,\kms$ \citep{Klypin2013}.
Results of the simulations for halos and subhalos are presented in
Figure~\ref{fig:LCDM}. At small $V$ the circular velocity function is
very close to a power-law. This power-law behaviour of the velocity
function is consistent with the results of much high resolution
simulations of individual halos and small regions
\citep[e.g.,][]{Diemand2008,Klypin2011,Sawala2014}. It allows us to
extrapolate our results to smaller values of $V$.

 We use the following approximations for the
differential circular velocity functions for halos and subhalos in the
range $V=(10-400)\,\kms$:

\begin{equation}
\frac{dN}{d\log_{10}V} =  A\left(\frac{V}{100\,\mbox{km\,s}^{-1}}
                            \right)^{-2.90} h^3\,\mbox{Mpc}^{-3},
\label{eq:VFlcdm}
\end{equation}
where the normalization $A$ is equal to
\begin{equation}
A = \left\{ \begin{array}{ll}
        0.130,  & \mbox{WMAP7},\\
        0.186,  & \mbox{Planck}.\end{array}  
\right. \label{eq:VFnorm}
\end{equation}

We use also predictions of the velocity function for the Warm Dark
Matter models made by \citet{Schneider2014} for models with thermal
neutrino masses $m_{\rm wdm}=1,2,4$~KeV. The velocity function was
derived from halo mass function \citep{Schneider2013} and halo
concentration-mass dependence \citep{Schneider2012}. The WDM mass
functions were estimated for the WMAP7 cosmology using $N$-body
simulations, and were approximated with analytical models. These
estimates are done only for distinct halos, and thus they do not include
subhalos. The fraction of satellites for given circular velocity $V$ is
relatively small. We account for the missing subhalos in the
\citet{Schneider2014} data by multiplying the abundance of distinct
halos by factor 1.25, which is the same fraction of subhalos as in the
$\LCDM$ model for circular velocities $V\lesssim 200\,\kms$. This is
a good approximation for circular velocities above $\sim 80\,\kms$,
because the effects of WDM are relatively small for these velocities and
for neutrino masses considered here ($m_{\rm wdm}\gtrsim 1\,$KeV). For
smaller velocities this likely overestimates the effect, however
there must be a significant number of small satellites to explain
dwarf satellites in the Local Group and in the Local Volume. So, our
estimate of the fraction of subhalos in the WDM models seems to be
reasonable.

\begin{figure}
\centering
\includegraphics[width=0.48\textwidth]{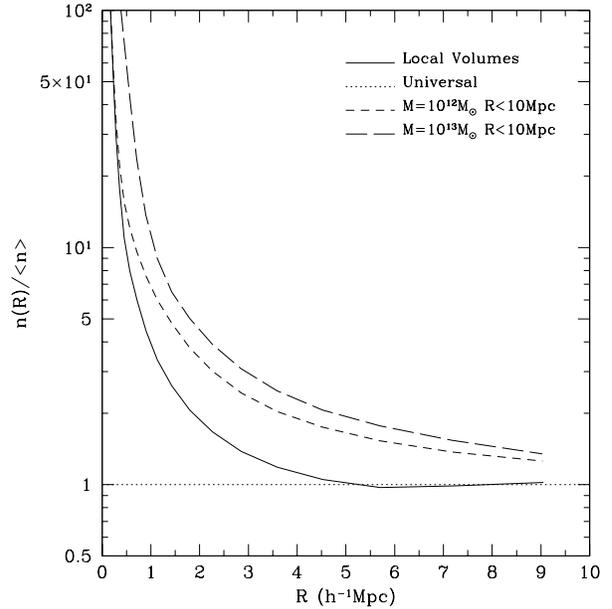}
\caption{ Number-density of halos and subhalos at the distance $R$
  from center of Local Volume candidates in the BolshoiP $\LCDM$
  cosmological simulation (full curve). The candidates are centered at
  halos with virial mass $(1-2)\times 10^{12}\Msunh$ and have from 6
  to 12 (sub)halos with $\Vmax > 170\,\kms$ inside sphere of radius of
  $7\Mpch$. The dashed and long-dashed curves show the number-density
  profiles for spheres centered on $10^{12}\Msunh$ and $10^{13}\Msunh$
  halos without the any constraints on the number of large halos in
  the region. The spike of the number-density at small ($<1\Mpch$)
  radii is the reflection large correlation function of halos at small
  distances. At larger radii the Local Volume candidates have the
  number-densities close to the average while unconstrained $7\Mpch$
  regions are on average significantly overdense.}
\label{fig:LVdens}
\end{figure}

\begin{figure}
\centering
\includegraphics[width=0.48\textwidth]{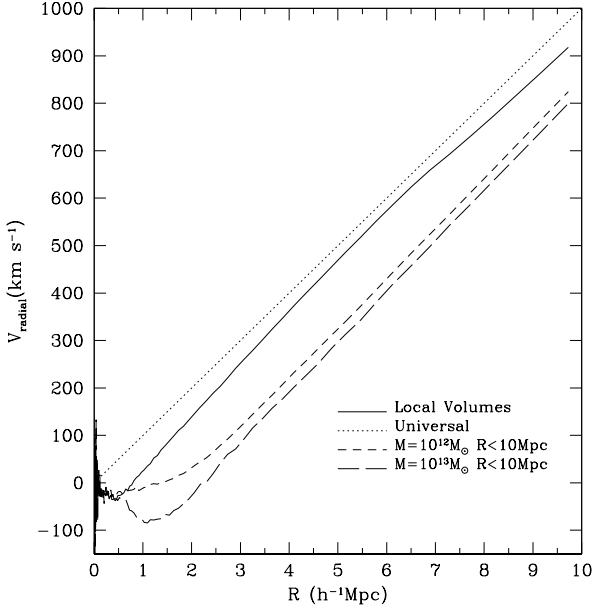}
\caption{ Average radial velocity of (sub)halos at distance $R$ from
  centers of Local Volume candidates (full curve). Velocities for
  unconstrained samples centered on $10^{12}\Msunh$ (short dash) and
  $10^{13}\Msunh$ (long dash) halos are also shown. The dotted line
  shows the Hubble velocity. The deviations from the Hubble flow are
  relatively small for the Local Volume candidates. Large average
  overdensities of unconstrained samples seen in
  Figure~\ref{fig:LVdens} result in slowing the expansion rates for
  these cases observed as large deviations from the Hubble flow even
  on large $\sim 5-10\,\Mpch$ distances.  }
\label{fig:LVvrad}
\end{figure}

\begin{figure}
\centering
\includegraphics[width=0.48\textwidth]{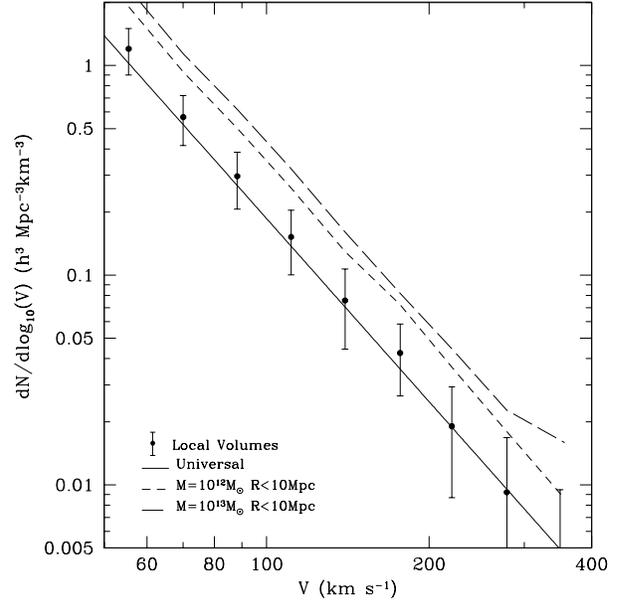}
\caption{ Velocity function of halos and subhalos for different samples
  in the BolshoiP cosmological simulation. The full curve show the
  velocity function for all (sub)halos in the simulation. Circles with
  errorbars show the average and rms deviations for Local Volume
  candidates. Unconstrained samples are shown with short and long
  dashed curves. On average, the Local Volume candidates show little
  systematic deviations from the velocity function in the whole
  simulation. The rms fluctuations are larger than the Poissonian
  estimates, but they are still reasonably small. }
\label{fig:LVvfunc}
\end{figure}

In practice we use an analytical approximation to the data provided by
\citet{Schneider2014}.  The following analytical form provides a 2\%
accurate fit the data on the WDM circular velocity function corrected
for the subhalos abundance:

\begin{equation}
\frac{dN}{d\log_{10}V}\left|\begin{array}{l} \\\!_{\rm WDM} \end{array} 
              \right.= \frac{1}{S(V,m_{\rm wdm})}  \frac{dN}{d\log_{10}V}
         \left|\begin{array}{l} \\\!_{\Lambda\rm CDM} \end{array} \right.,
\label{eq:VFwdm}
\end{equation}
where the WDM suppression factor $S(V,m_{\rm wdm})$ is 
\begin{eqnarray}
 S(V,m_{\rm wdm}) &=& 1 +7200\left[1 +\left(\frac{m_{\rm wdm}V}{23\,\mbox{km\,s}^{-1}}
                      \right)^{6.2}\right] \times \nonumber\\
       &&\phantom{mmmm} \left(\frac{m_{\rm wdm}V}{10\,\mbox{km\,s}^{-1}}\right)^{-8.2}.
\label{eq:Swdm}
\end{eqnarray}
Here the WDM mass $m_{\rm wdm}$ is given in the units of 1~KeV.
Figure~\ref{fig:LCDM} presents results for velocity function in the
WDM models.  We note that the WDM velocity function has a simple
dependence on mass $m_{\rm wdm}$: the suppression factor $S(V,m_{\rm
  wdm})$ in eqs.~(\ref{eq:VFwdm}-\ref{eq:Swdm}) depends only on the
product $Vm_{\rm wdm}$. This is likely related to the fact that the
$\LCDM$ velocity function for the relevant velocity range $V\lesssim
200\,\mbox{km\,s}^{-1}$ is nearly a power-law, and, thus, it is
scale-free. The only scale dependence comes from the power spectrum
suppression due to $m_{\rm wdm}$.

This simple scaling relation of the WDM models allows one to estimate
the WDM predictions for any $m_{\rm wdm}$ and for different
cosmological parameters. When comparing with observations, we re-scale
the abundance of halos in the WDM models to the Planck cosmological
parameters using eqs.~(\ref{eq:VFlcdm}-\ref{eq:VFnorm}) for $dN/d\log
V|_{\Lambda\rm CDM}$ and the suppression factor $S$ given by
eq.~(\ref{eq:Swdm}).

Detailed analysis of the effects of baryons on the circular velocity
function was done by \citet{Trujillo2011} and \citet{Dutton2011}.  The
impact of baryons is small for galaxies hosted by halos with
$V\lesssim 100\,\kms$. More massive galaxies are more affected. For
example, modeling of the Milky Way galaxy \citet{Klypin02}, which used
the $\LCDM$ predictions (the NFW profile with realistic concentration
and spin parameter), indicates that the dark-matter-only maximum of
the circular velocity should be $\sim (160-170)\,\kms$. Taking
$V\approx 230\,\kms$ for circular velocity of the Milky Way, we find
that the baryons increase the circular velocity by a factor
$1.3-1.4$. In this paper we approximate the complex effects of baryons
studied by \citet{Trujillo2011} using a simple fitting function of the
results presented in Figure~12 of \citet{Trujillo2011}
\footnote{Unlike \citet{Trujillo2011}, who estimate circular
  velocities at radius of 10~kpc, we always use the the maximum
  circular velocity $\Vmax$}.  Specifically, we use the following fit:

\begin{eqnarray}
V &=& V_{\rm DM+bar}\left[1+0.35x^6\left(1+x^6\right)^{-1}\right]^{-1}, \\
        x&\equiv& \frac{V_{\rm DM+bar}}{120\,\mbox{km\,s}^{-1}},
\end{eqnarray}
where $V$ is maximum of circular velocity for
dark-matter-only predictions, and $V_{\rm DM+bar}$ is the the circular
velocity corrected for the effect of baryons.

\section{How typical is the Local Volume: theoretical view}

The Local Volume has many hundreds of galaxies, and, thus one expects
that to some degree it is a representative sample of galaxies inside a 
$\sim 10$~Mpc sphere selected not to be centered on a void or a cluster
of galaxies. Indeed, the B-band luminosity function presented in
Figure~\ref{fig:LumFun} is consistent with this expectation: the
luminosity function in the Local Volume is close to the luminosity
function in the SDSS catalog for an  overlapping range of magnitudes. How
 can this be considering that the size of the Local Volume is relatively small?

 One needs to keep in mind that the Local Volume is not a randomly
 selected sphere of 10~Mpc centered on a Milky-Way type galaxy. If it
 were, the region would have been a factor of $\sim 2-3$ overdense
 \cite[e.g.,][]{Hogg2005,Zehavi2011} reflecting a large amplitude of
 the galaxy-galaxy correlation function on a few Megaparsec scale. It
 also would have had a very large level of fluctuations because of
 cosmic variance: occasionally it would fall on either a cluster of
 galaxies or a nearly empty void.  One property of the Local Volume
 makes it more representative of the average Universe, and makes the
 density inside the 10~Mpc sphere close to the average density in the
 Universe. This is the fact that the Local Volume does not have large
 groups or clusters inside its boundaries.

We can measure the magnitude of fluctuations and test the effects of
different selection conditions by studying properties of analogs of
the Local Volume in numerical simulations of the $\LCDM$~model. For
this purpose we use the BolshoiP simulation. As a center of a Local
Volume candidate we chose a distinct halo in the mass range
$(1-2)\times 10^{12}\Msunh$. We then select only those candidates that
have a number of large halos and subhalos $N_{\rm large}=6-12$ with
the maximum circular velocity larger than $\Vmax > 170\,\kms$ within a
radius of $7\Mpch$. This number of large halos is compatible with the
number of large galaxies observed in the Local Volume (see, for
example, Figure~\ref{fig:VDdiagrams}). We use halos and subhalos with
maximum circular velocity larger than $\Vmax>50\,\kms$ without any corrections for baryons.

\begin{figure*}
\centering
\includegraphics[width=0.48\textwidth]{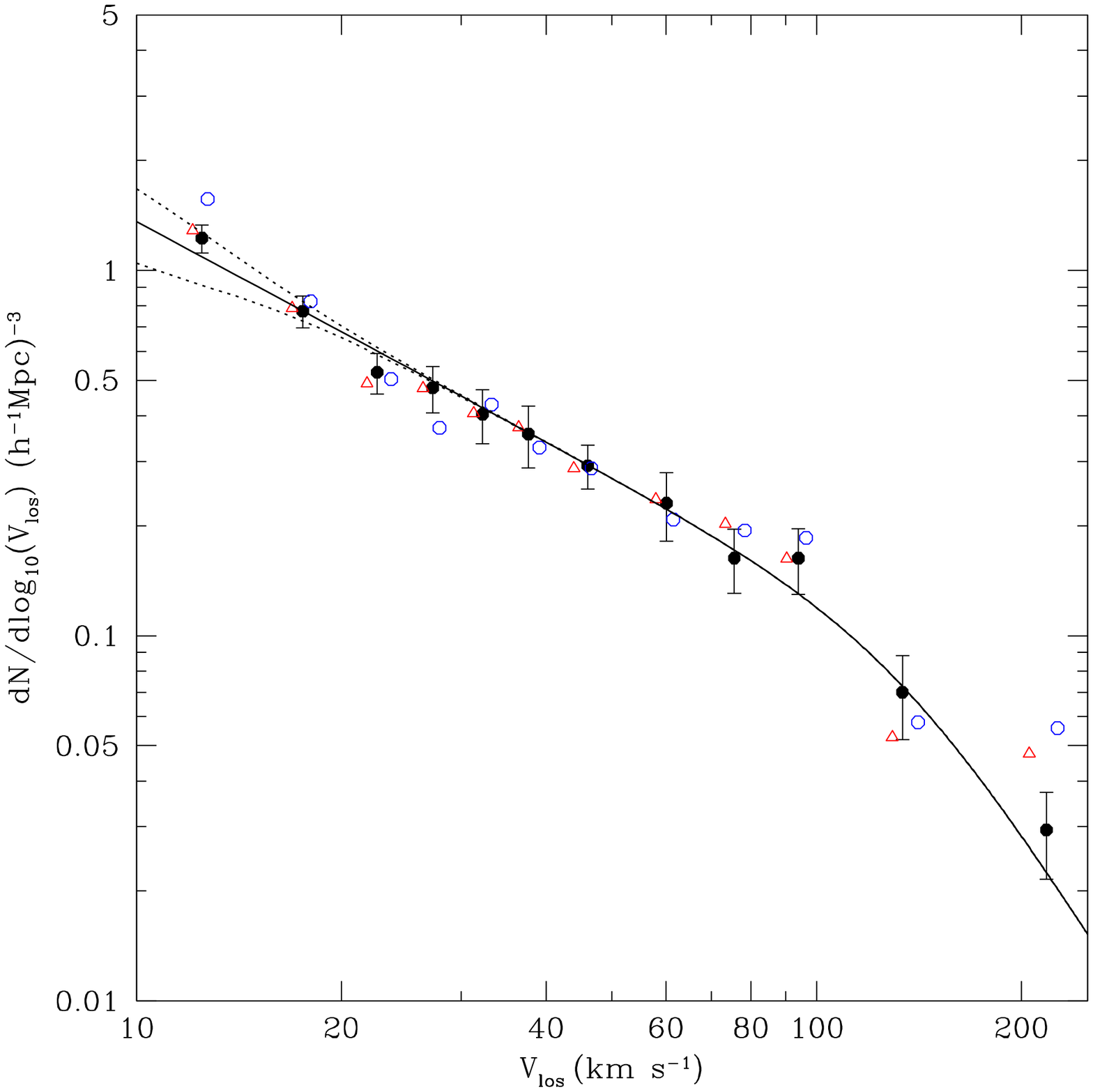}
\includegraphics[width=0.48\textwidth]{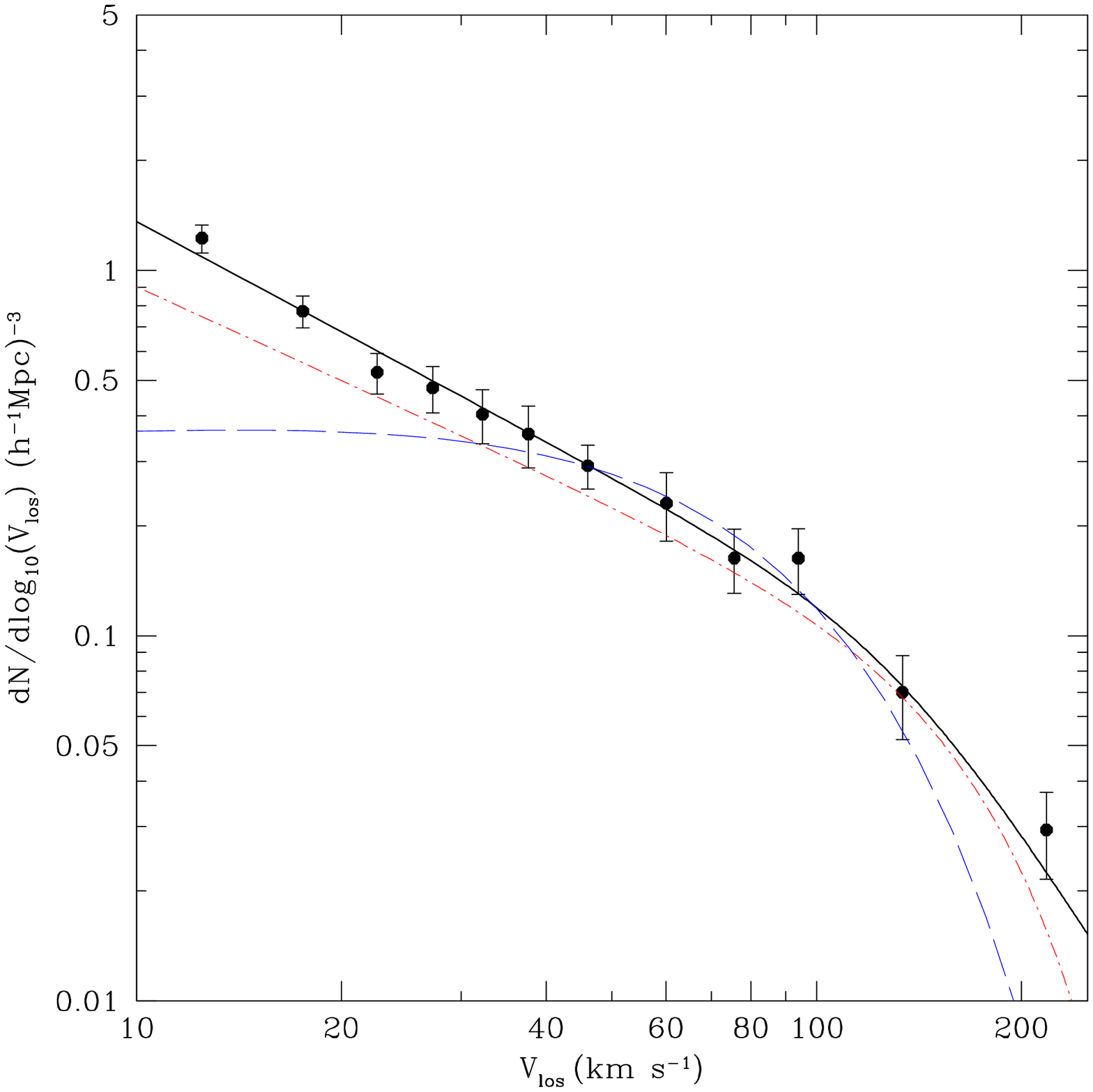}
\caption{ Distribution function of line-widths $V_{\rm los}$ for
  galaxies in the Local Volume. {\it Left:} Different symbols show
  results for different subsamples of the Local Volume.  Filled
  symbols with error bars present results for the 10~Mpc sample, while
  the open circles and triangles are for 6~Mpc and 8~Mpc subsamples
  correspondingly. Error bars show poissonian statistical
  fluctuations. The full curve presents the analytical fit given by
  eq.~(\ref{eq:fit}).  Comparison of the subsamples indicates
  stability of the velocity function for variations in the sample
  size. Dotted curves show effect of 30\% uncertainty in the selection function eq.(4).
 {\it Right:} Comparison of the distribution function of
  line-widths for galaxies in the Local Volume (circles with error
  bars and the full curve) with \HI~ half line-width measurements in
  ALFALFA \citep[dot-dashed curve,][]{Papastergis2011} and HIPASS
  \citep[long-dashed curve,][]{Zwaan2010} surveys. In addition to
  gas-rich late-type galaxies the Local Volume sample has early-type
  galaxies, which are  missed in the \HI~ surveys.}
\label{fig:VFproj}
\end{figure*}

The average number-density profile and the average radial velocity of
(sub)halos in the Local Volume candidates are shown in
Figures~\ref{fig:LVdens} and \ref{fig:LVvrad}.  As one may have
expected, there are substantial overdensities and, consequently large
deviations from the Hubble flow for scales below $\sim 2\,\Mpch$. This
is just the reflection of the fact that the sample is centered on a
large halo and that halos highly correlate on small scales. However,
the Local Volume candidates have dramatically smaller deviations on
larger scales, which is a reflection of the selection condition; just
as the real Local Volume, these candidates are not allowed to have
large groups or clusters inside their boundaries. 

Figure~\ref{fig:LVvfunc} shows the velocity functions for the Local
Volume candidates inside spheres of $7\Mpch$ radius as well as for two
other unconstrained samples.  On average, the Local Volume candidates
show little systematic deviations from the velocity function of the
whole simulation. The rms fluctuations are larger than the Poissonian
estimates, but they are still reasonably small.

\begin{figure*}
\centering
\includegraphics[width=0.48\textwidth]{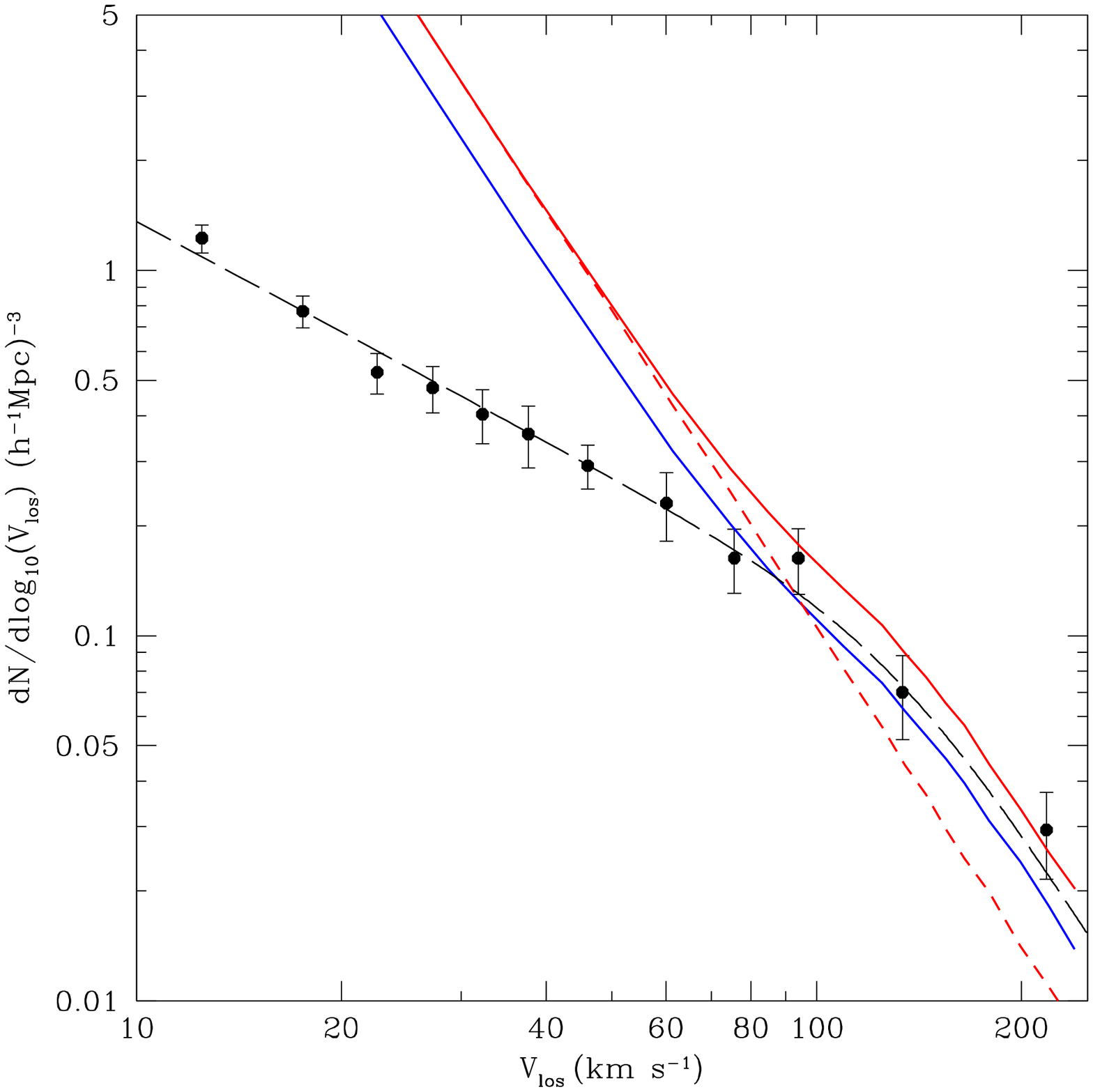}
\includegraphics[width=0.48\textwidth]{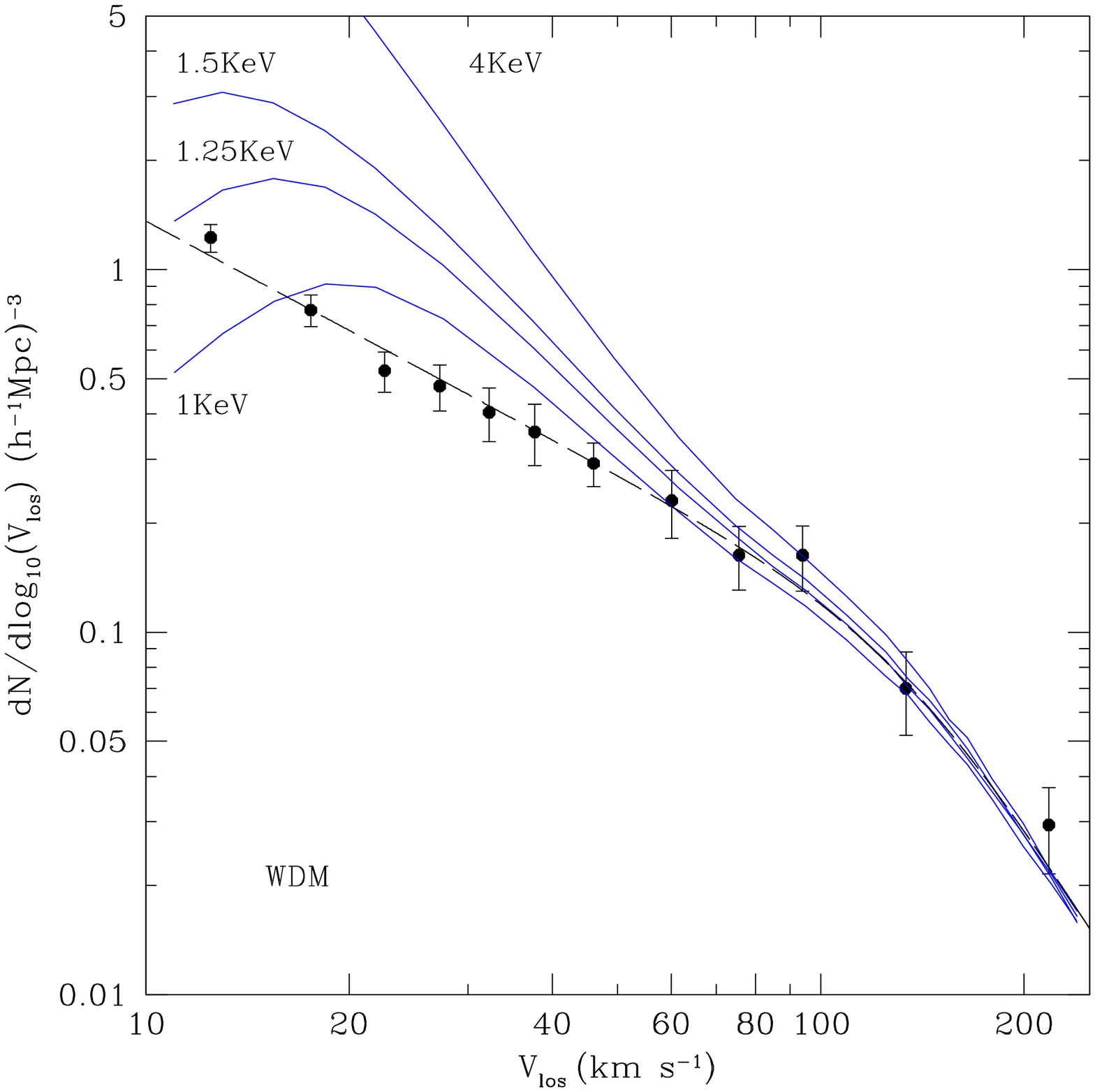}
\caption{ Comparison of the distribution function of line-widths
  $V_{\rm los}$ for galaxies in the Local Volume with theoretical
  predictions for the LCDM (left panel) and the Warm Dark Matter
  models (right panel, Planck cosmological parameters). 
  Filled circles and the long-dashed curve present velocity function
  for the 10~Mpc sample. {\it Left:} Theoretical predictions for the \LCDM~ model
  with the Planck cosmological parameters are presented by the upper
  full curve. The lower full curve shows predictions for the WMAP7
  cosmology. The short-dashed curve shows the predictions of the dark
  matter-only estimates without correction for baryon infall. Enhanced
  mass of baryons (mostly due to stars) in the central halo regions
  results in the increase of the circular velocity observed in this
  plot as the shift from the dashed to the full curve.}
\label{fig:VFtheory}
\end{figure*}

\section{Results: Velocity function of galaxies}

The left panel in Figure~\ref{fig:VFproj} shows estimates of the galaxy
velocity function in the Local Volume for different subsamples. The
number-density in each subsample is not the same mostly because of a
large excess of bright galaxies at 3.5-4~Mpc distance from the Milky
Way. Effect of the fluctuation declines with the volume, and for the
10~Mpc sample the number-density is close to the average
number-density in the much larger SDSS sample, as indicated in
Figure~\ref{fig:LumFun}. For this reason estimates of the abundance of
galaxies in the 6~ and 8~Mpc samples were normalized to have the same
number-density of galaxies brighter than $M_B=-14$ as in the 10~Mpc
sample.  Once normalized to the same number-density, comparison of
different subsample serves as indicator of stability of estimates of
the velocity function.

The full curve in the plot shows a fit to the 10~Mpc data:
\begin{equation}
  \frac{dN}{dlog_{10}V_{\rm los}} = 13.6V_{\rm los}^{-1}\left[1+\left(\frac{V_{\rm los}}{135\,{\kms}}
                           \right)^4\right]^{-1/2}
  h^{3}\,{\rm Mpc}^{-3}
\label{eq:fit}
\end{equation} 
Errors of this approximation are dominated by the cosmic varience,
which was discussed in Section~4 and presented in
figure~\ref{fig:LVvfunc}. The errors are $\sim 20\%$ for circular
velocities $V < 100\kms$.

In the right panel of Figure~\ref{fig:VFproj} we compare our estimates
with the HIPASS \citep{Zwaan2010} and ALFALFA \citep{Papastergis2011}
results. Our estimates are systematically larger because HIPASS and
ALFAFA do not include gas poor early-type galaxies while those
galaxies are included in the Local Volume sample. However, the
fraction of the early-type galaxies is relatively small, as indicated
in  the top panel in Figure~\ref{fig:LumFun}. So, the agreement
between different catalogs is reasonably good for velocities in the
range $V_{\rm los} \approx (25-150)\,\mbox{km\,s}^{-1}$. At smaller
velocities the Local Volume results are substantially above HIPASS and
ALFALFA mostly because the fraction of gas-poor galaxies increases and
likely because of incompleteness in the HIPASS data.

We compare our results with theoretical predictions in
Figure~\ref{fig:VFtheory}.  In the left panel we confront
observational results with the predictions of the $\LCDM$ model with
the Planck parameters. Correction for baryon infall becomes
progressively more important for $V_{\rm los} \gtrsim
60\,\mbox{km\,s}^{-1}$.  Thanks to this correction, the $\LCDM$
model makes a reasonably accurate account for the abundance of bright
galaxies. At smaller velocities $V_{\rm los}$ effects of baryons are not
significant. Here the theory clearly has substantial problems.

However, the real problem for the model is the abundance of relatively
large dwarf galaxies with $V_{\rm los} = (30-40)\,\mbox{km\,s}^{-1}$.
The $\LCDM$ model overpredicts the abundance of those galaxies by a
factor of 3.5-7.5 for the model with the Planck parameters.
 These
galaxies are hardly affected by possible effects of reionization and
cannot be stopped from forming stars by few supernovae. These galaxies
are relatively bright with $M_B\approx -15-16$. For these luminosities
the Local Volume catalog is nearly complete. We definitely can exclude
the possibility that $\sim (70-80)\%$ of these galaxies are missed. In
short, it is difficult to reconcile the $\LCDM$ predictions with
observations.

The WDM models are somewhat better. In the right panel of
Figure~\ref{fig:VFtheory} we test the WDM models. Models with
$m_{\rm wdm}> 1.5$~KeV can be excluded because they do not provide
enough reduction of the number of dwarf galaxies.  Smaller masses help
to suppress the low-mass tail of the velocity function, but the shape
of $dN/d\log V$ is not right. For example, the model with
$m_{\rm wdm}= 1.0$~KeV is still above data points with
$V_{\rm los} = (20-40)\,\mbox{km\,s}^{-1}$, and it misses the smaller
$V_{\rm los}$'s. Further decreasing $m_{\rm wdm}$ would improve
points with $(20-40)\,\mbox{km\,s}^{-1}$ but ruin $~10\,\kms$.

The circular velocity function  $dN/d\log V$ of observed galaxies can be
reconstructed in a statistical sense by following the same steps,
which we do when we make the transition from theoretical circular
velocity function  $dN/d\log V$ to the distribution of line-width $dN/d\log V_{\rm los}$. In the
case of observations we assume an analytical function with free
parameters. For each set of the parameters we make a prediction for
$dN/d\log V_{\rm los}$ by assuming a random orientation of disk
galaxies and by taking the observed fraction of early type
galaxies. We then change the free parameters until we  get an acceptable fit
to the data. The best fit to observed circular velocity function of galaxies is:
\begin{eqnarray} 
\frac{dN}{d\log_{10}V} &=& 0.18\left[ \frac{V}{100\,\mbox{km\,s}^{-1}}\right]^{-1}\times\nonumber \\
        &&     \exp\left(-\left[ \frac{V}{250\,\mbox{km\,s}^{-1}} \right]^3\right) h^3\,\mbox{Mpc}^{-3}.
\label{eq:VFfit}
\end{eqnarray}

\begin{figure}
\centering
\includegraphics[width=0.46\textwidth]{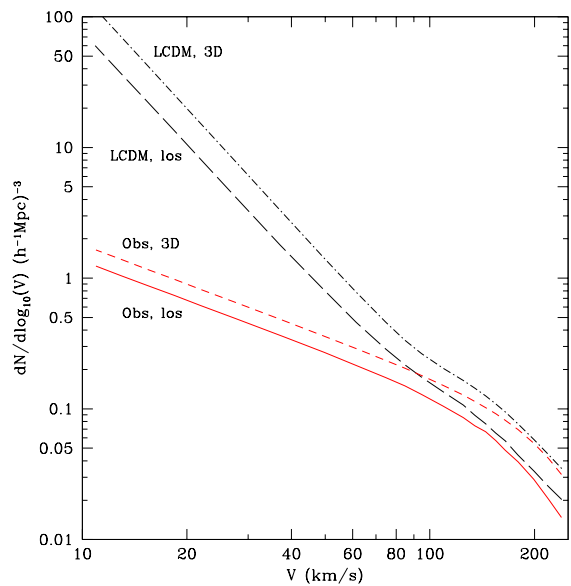}
\caption{Relation between 3D circular velocity $dN/d\log V$ and
  line-width $dN/d\log V_{\rm los}$ functions for observations (short
  dashed and full curves) and for the $\LCDM$-Planck model (dot-dashed
  and long dashed curves).  The short-dashed curve shows our estimate
  of the circular velocity function in the Local Volume.  It produces
  the distribution of line-widths that accurately fits the
  observations.  The disagreement between the $\LCDM$ model and
  observations becomes slightly worse for the 3D circular velocities
  as compared with the line-of-sight line width.}
\label{fig:Corrected}
\end{figure}

Note that the slope of the circular velocity function $dN/d\log V$ is
close, but slightly smaller that the slope of the related line-width
function $dN/d\log V_{\rm los}$. One can show analytically that a pure
power-law $dN/d\log V$ gives a power-law line-width function
$dN/d\log V_{\rm los}$ with exactly the same slope. The small change in the
slope between $dN/d\log V$ and $dN/d\log V_{\rm los}$ seen in
Figure~\ref{fig:Corrected} is, thus, due to the bending of $dN/d\log V$ at
large $V$.

Comparison of the observed and theoretical circular velocity functions
presented in Figure~\ref{fig:Corrected} leads to the same conclusion,
which we found comparing the line-width functions in
Figure~\ref{fig:VFtheory}: the $\LCDM$ model gives a good fit for
massive galaxies with $V_{\rm los}>70\,\mbox{km\,s}^{-1}$, but it
has problems explaining the abundance of galaxies with small
velocities. However, the disagreement is slightly worse for $dN/d\log
V$ functions as compared with $dN/d\log V_{\rm los}$. For example, at
$V_{\rm los}=40\,\mbox{km\,s}^{-1}$ the disagreement with the $\LCDM$-Planck
model was factor 3.5 for the line-width functions. It is factor of 6
for the circular velocities. Qualitatively, it is clear why the
disagreement is worse in $V$-space: some fraction of galaxies with given
$V_{\rm los}$ are intrinsically larger galaxies with large $V$, for with
the $\LCDM$ predicts correct abundance.

\section{Discussion}

 Estimates of the abundance of galaxies with a
given line-width $V_{\rm los}$ presented in Figure~\ref{fig:VFproj}
for different observational samples shows that results mostly agree
for intermediate-size galaxies with $V_{\rm los} \approx
(25-150)\,\mbox{km\,s}^{-1}$. The Local Volume results are
systematically above the HIPASS \citep{Zwaan2010} and ALFALFA
\citep{Papastergis2011} estimates, but this is mostly due to the fact
that \HI~ measurements do not cover early-type galaxies, which are
present in the Local Volume. The agreement between the ALFALFA
\citep{Papastergis2011} and Local Volume results is particular good
once ALFALFA results are corrected by the fraction of early-type
galaxies in the Local Volume as presented in Figure~\ref{fig:LumFun}.
This is very encouraging because it indicates that we finally have an
accurate measurement of the abundance of galaxies in a broad range of
velocities $(10-200)\,\mbox{km\,s}^{-1}$.

\begin{figure}
\centering
\includegraphics[width=0.48\textwidth]{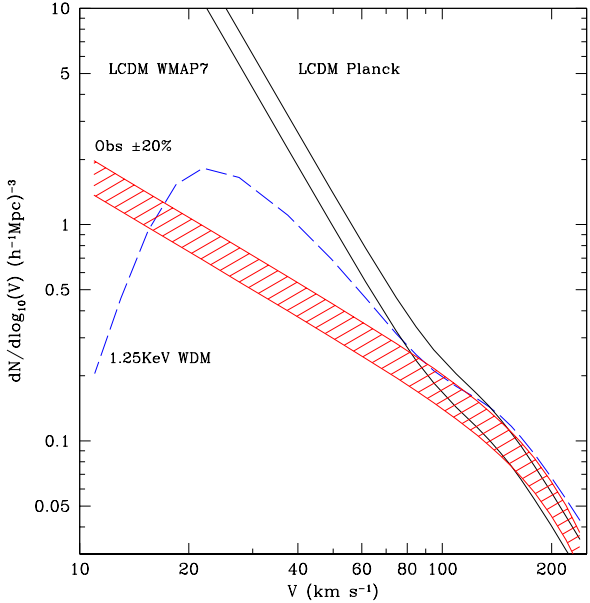}
\caption{Comparison of the observed and theoretical estimates of the
  circular velocity function of galaxies. The shaded area shows the
  region of the observed velocity function -- a strip of $\pm$20\%
  around the reconstructed velocity function in the Local Volume as
  given by eq.~(\ref{eq:VFfit}). The full curve shows the $\LCDM$
  model predictions. The dashed curves shows the best prediction for
  the WDM model with thermal neutrino mass of $m_{\rm
    dwm}=1.35$~KeV. Both theoretical models have severe problems. The
  WDM model predicts a wrong shape for the velocity function, it fails
  by a factor of 2-3 at small velocities while still overpredicting the
  abundance of $30\,\mbox{km\,s}^{-1}$ galaxies. The $\LCDM$-Planck model
  overpredicts the abundance of dwarf galaxies with $V\lesssim 60\,\mbox{km\,s}^{-1}$.}
\label{fig:VFtrue}
\end{figure}


Figure~\ref{fig:VFtrue} compares observational estimates for the
circular velocity function of galaxies with theoretical predictions
for the $\LCDM$ and WDM models. As one may have expected, the $\LCDM$
model dramatically overpredicts the abundance of dwarf galaxies. The
WDM is not better.  It was designed to fix the small-scale problems of
the $\LCDM$. As far as the abundance of dwarfs is concerned, the WDM
still fails. 

 As Figure~\ref{fig:VFtrue} shows, the WDM
model with $m_{wdm}=1.25$~KeV predicts a factor $\sim 2-3$ too few
dwarfs with $V=(10-15)\,\mbox{km\,s}^{-1}$ and a factor of 2 too many
at $V\approx 25\,\mbox{km\,s}^{-1}$. As plots in
Figure~\ref{fig:VFtheory} indicate, decreasing neutrino mass to
$m_{wdm}=1.0$~KeV fixes the problem with
$V\approx 25\,\mbox{km\,s}^{-1}$, but it also ruins the small-scale
tail by reducing the abundance of $V=(10-15)\,\mbox{km\,s}^{-1}$
dwarfs by another factor of 2. Thus, {\it there seem to be no neutrino
  mass that produces an acceptable fit to the data}.

Additional limitations on the WDM model are coming from the
clustering observed in the Lyman-$\alpha$ forest
\citep[e.g.,][]{Seljak2006,Viel2008,Viel2013}. Recent results of
\citet{Viel2013} constrain the (thermal) neutrino mass $m_{wdm}>2$~KeV
at $4\sigma$ level. Our results indicate that neutrino mass above
2~KeV is incompatible with the observed abundance of field dwarf
galaxies in the Local Volume.  Our conclusions regrading the inability
of WDM models to explain the observational data are in
agreement with those of \citet{Schneider2014}.

 In spite of the fact that our estimates
of the abundance of field galaxies are above the previous estimates,
the $\LCDM$ predictions for the velocity function are still well above
the observations. This is neither new
\citep{Tikhonov09,Trujillo2011,Schneider2014} nor surprising. The overabundance
of satellites \citep{Klypin1999,Moore1999} in the $\LCDM$ model is a
well established problem, and it has the same origin as the
overabundance of field galaxies. However, the galaxy velocity function
rises the problem to a different level.

It is interesting to compare the overabundance of satellites in the
Local Group with the overabundance of field galaxies:

\noindent $\bullet$ Most of the galaxies in the Local Volume are not
satellites (though there are some). Thus, the problem with the field
galaxies cannot be solved by appealing to effects such as the tidal
forces and ram pressure stripping due to the central ``parent''. For
satellites close to their parent (say, distances less than virial
radius) the stripping and tides are a possible solution
\citep{Zolotov2012,Arraki2014} for the problems with structural
properties of the largest dwarf spheroidal galaxies -- the ``too big
to fail'' problem \citep{Boylan-Kolchin2011}.

However, these processes are not expected to be efficient enough for
satellites in the outskirts ($\gtrsim 300\,$kpc) of the Local
Group. Indeed, \citet{Garrison2014} find an excess of theoretically
predicted large ($\Vmax>30\,\kms$) satellites. This situation is
similar to the problem of the abundance of field dwarf galaxies, which
we find in this paper.

\noindent $\bullet$  The main problem in the Local Group was related with dwarf
  spheroidal galaxies. In the Local Volume most of ``problematic''
  galaxies are dwarf irregular galaxies, which are star-forming
  gas-reach galaxies.

  \noindent $\bullet$ The satellite problem starts at relatively small
  galaxies with $V\lesssim 20\,\mbox{km\,s}^{-1}$ and
  $M_{\rm vir} \lesssim 10^{9}\Msunh$.  There is practically no issue
  with the number of satellites with $V> 30\,\mbox{km\,s}^{-1}$: the
  theory predicts as many as observed
  \citep[e.g.,][]{Klypin1999,Kravtsov2010}. The situation is much
  worse in the field, where the disagreement is already very severe
  for galaxies with $V = 40\,\mbox{km\,s}^{-1}$ and virial masses
  $M_{\rm vir} \approx 10^{10}\Msunh$. This can be expressed in a
  number of ways. At $V = 40\,\mbox{km\,s}^{-1}$ the ratio of
  $dN/d\log V$ of the predicted (eq.~(\ref{eq:VFlcdm})) to the
  observed (eq.~(\ref{eq:VFfit})) number of galaxies is 6 for the
  Planck cosmology. The total number of galaxies in the Local Volume
  (distances $<10$~Mpc) with circular velocities in the range
  $V = (30-50)\,\mbox{km\,s}^{-1}$ is $\sim 200$, while the theory
  predicts $\sim 1000$.

\medskip
What can possibly be a solution for the field problem? 
\medskip

 {\it Observations:} It is possible and very likely that a number of
small dwarfs galaxies with $V<20\,\mbox{km\,s}^{-1}$ were
missed. However, at $V =20\,\mbox{km\,s}^{-1}$ the disagreement with
the theory is a factor of 20. In order to reconcile observations with
the $\LCDM$ model, most of the dwarfs must have been missed in the
Local Volume sample: an unlikely proposition considering the convergence
of results on the luminosity function at $M_B=-14$.

The main problem is with larger galaxies in the range of velocities $V
= (30-50)\,\mbox{km\,s}^{-1}$. These galaxies are bright,
$M_B\approx -16$, and it is unrealistic to assume that the Local
Volume sample missed 1000 of them. The only remote
possibility is that a large fraction of the galaxies are low surface
brightness dwarf spheroidal galaxies with surface brightness well below
25 mag per square arcsecond. So far, none of these bright and extra
low surface brightness galaxies have been found.

{\it Theory:} It is difficult to resolve the overabundance of field
galaxies because some of the problematic galaxies are relatively large
with $M_{\rm vir} \approx 10^{10}\Msunh$ and $\Vmax \approx 30\,\kms$. For
example, photoheating during reionization hardly can affect these
objects. Another possible way out is not to have the star formation in most of
these galaxies. However, this does not help either because these objects would
keep their neutral hydrogen, and thus would be observed as \HI~ clouds
without stellar light. However, those massive dark \HI~clouds have not
been found. Mass of the clouds should be $M_{HI}\sim (3-5)10^7\Msun$,
if they follow the Baryonic Tully-Fisher relation.

One may think that flattening of the dark matter cusp in dwarf
galaxies in numerous episods of contraction and expansion due to
centrally concentrated bursts of star formation
\citep[e.g.,][]{Mashchenko2006,Pontzen2012} may solve the problem with
the abundance of field dwarfs. Indeed, large variations in the
gravitational potential of a forming dwarf galaxy are capable of
reducing the dark matter density in the central galaxy region, which
in turn reduces the circular velocity
\citep{Governato2010,Teyssier2013,DiCintio2014,Madau2014}.
Unfortunately, this reduction in the central dark matter density does
not automatically resolve the issue.

The dark matter, which was heated and pushed from the cusp, does not leave the
galaxy. It just moves away from the central cusp and settles at
somewhat larger radius \citep{Teyssier2013,Madau2014}. Within this
radius the total mass does not change, and thus the circular velocity
does not change either.  The question is how large is the radius and
how does it compare to the radius that defines the observed \HI~line width.

Rotation curves in dwarf galaxies are rising even at large
radii. Thus, it is the outskirts of galaxies that define the observed
\HI~line width \citep[e.g.,][]{Oh2008,Moiseev2014}. The problem is
that the neutral hydrogen in galaxies extends to very large radii:
substantially larger than the optical radius, where thrashing and
heating of the dark matter happens. For example, data in
Figure~\ref{fig:GasExtent} indicates that gas in galaxies with
rotational velocity $(30-40)\,\kms$ extends to 2--4~kpc. (See also
Figure~3 in \citet{Ferrero} and
Figures~5 and B.1 in \citet{Papastergis2014}  for simular
results). Hydrodynamical simulations show that the size of the core in
such galaxies should be $\sim 1\,\kpc$
\citep{Teyssier2013,DiCintio2014,Madau2014}, which is not enough to
substantially reduce \HI~line width at $\sim 4\,\kpc$. The problem is
even worse for smaller galaxies with rotational velocity
$\sim 20\,\kms$ and virial mass $\Mvir \approx 10^9\Msun$. Current
numerical simulations of these galaxies \citep{DiCintio2014,Madau2014}
indicate that there is not enough energy in the stellar feedback to
flatten dark matter cores of these dwarfs.

It is possible that current simulations may not implemented all
effects of the stellar feedback, and that better models will show
larger decline in the density.  Another possible solution for the
overabundance problem is the possibility that in reality there are
many missed very low surface brightness galaxies with just enough star
formation to keep the gas ionized. So, the galaxies are not detected in
both \HI~ and optical observations.

After our paper was submitted for publication, \citet{Papastergis2014}
came to similar conclusions the regarding the extend of the HI gas in
dwarf galaxies: neutral gas extends far enough to probe the part of
the circular velocity curve that defines value of $\Vmax$.  They also
confirm our findings on the disagreement between observations and
theory on the abundance of galaxies with given circular velocities.
 
\section{Conclusions}

The abundance of galaxies as a function of their circular velocity
$dN/dV$ is a fundamental statistics, which provides a sensitive probe
for theoretical predictions
\citep{Cole1989,Shimasaku1993,Gonzalez2000,
  Zavala2009,Trujillo2011,Schneider2014}. Because it is difficult to
measure the velocity function, only recently observations became capable of
producing reasonably converging estimates of $dN/dV$ for different
samples \citep{Zwaan2010,Papastergis2011}.  
Abundance of galaxies with given circular velocity $V$ or line-width
$V_{\rm los}$ in the Local Volume (distances less than 10~Mpc) provide
a valuable information that can be difficult to obtain using other
samples \citep{Karachentsev2004,Karachentsev2013} because in the Local
Volume we observe small galaxies, and the sample has galaxies of all
morphological types.

We find that the observed velocity function of galaxies of all
morphological types has a shallow slope
$dN/d\log V \propto V^{\alpha}, \alpha \approx -1$ at small velocities
and a relatively steep decline at large velocities. Equations (11) and (12)
give approximations for the observed line-width $V_{\rm los}$ and for
the reconstructed 3D circular velocity functions $V$.

Results for $dN/d\log V_{\rm los}$ in the Local Volume are consistent
with the velocity function in the much larger HI-based ALFALFA sample
\citep{Papastergis2011} once the latter is corrected for the
$\sim 10-20\%$ fraction of early-type galaxies not detected in HI
observations. This is important because it implies that we now have
$\sim 10\%$ accurate measurement of the galaxy velocity function for
all types of galaxies for circular velocities in the range
$V= (10-200)\,\kms$.

When comparing the Local Volume observational results with theoretical
predictions we find that the standard \LCDM\ model predicts abundances
of intermediate-size galaxies with $V_{\rm los}\gtrsim 70\,\kms$ and
corresponding virial masses $M_{\rm vir}\gtrsim 5\times 10^{10}\Msun$
that are in agreement with the observed abundance of galaxies.
However, the shallow slope and the normalization of the velocity
function for dwarf galaxies with $V_{\rm los}\lesssim 40\,\kms$
strongly disagree with the standard \LCDM\ model that predicts the
slope $\alpha =-3$ for dark matter halos and subhalos found in
$N-$body cosmological simulations. The Warm Dark Matter (WDM) models
cannot explain the data either, regardless of mass of WDM particle.

The overabundance problem of the field galaxies in many respects is different
from the more familiar overabundance of satellite galaxies in the
Local Group. Unlike the Local Group, where the problem is in the
abundance of gas-poor dSphs with $V\sim 10\,\kms$ and at small
distances $\lesssim 1\,\kpc$, in the field the problem is for gas-rich
star-forming galaxies with velocities as large as $V\sim 30-40\,\kms$
and on large distances $\gtrsim 2\,\kpc$.
\section*{Acknowledgments}

We are grateful to A.~Maccio and A.~Schneider for providing us tables
with the WDM velocity functions. We thank M.~Papastergis 
 for useful discussions and comments.  We acknowledge the
support of the NSF grants to NMSU, and the support of the grant
14-12-00965 from the Russian Science Foundation. O.N. thanks the
non-profit Dmitry Zimin's Dynasty Foundation for the financial
support. The Bolshoi and BolshoiP simulations were run on the Pleiades
supercomputer at the NASA Ames Research Center.

\label{lastpage}

\end{document}